\documentclass{article}

    \PassOptionsToPackage{numbers, compress}{natbib}

\usepackage{aimc2022}

\usepackage[utf8]{inputenc} 
\usepackage[T1]{fontenc}    
\usepackage{hyperref}       
\usepackage{url}            
\usepackage{booktabs}       
\usepackage{amsfonts}       
\usepackage{amsmath}       
\usepackage{nicefrac}       
\usepackage{microtype}      
\usepackage{graphicx}       
\usepackage{siunitx}
\usepackage{listings}       

\title{Notochord: a Flexible Probabilistic Model for Real-Time MIDI Performance}

\author{%
  Victor Shepardson\\
  Intelligent Instruments Lab\\
  Iceland University of the Arts\\
  Reykjavík, Iceland \\
  \texttt{victor@lhi.is} \\
  \And
  Jack Armitage \\
  Intelligent Instruments Lab\\
  Iceland University of the Arts\\
  Reykjavík, Iceland \\
  \texttt{jack@lhi.is} \\
  \And
  Thor Magnusson \\
  Intelligent Instruments Lab\\
  Iceland University of the Arts\\
  Reykjavík, Iceland \\
  \texttt{thor.magnusson@lhi.is} \\
}

\begin{document}

\maketitle

\begin{abstract}
Deep learning-based probabilistic models of musical data are producing increasingly realistic results and promise to enter creative workflows of many kinds. Yet they have been little-studied in a performance setting, where the results of user actions typically ought to feel instantaneous. To enable such study, we designed Notochord, a deep probabilistic model for sequences of structured events, and trained an instance of it on the Lakh MIDI dataset. Our probabilistic formulation allows interpretable interventions at a sub-event level, which enables one model to act as a backbone for diverse interactive musical functions including steerable generation, harmonization, machine improvisation, and likelihood-based interfaces. Notochord can generate polyphonic and multi-track MIDI, and respond to inputs with latency below ten milliseconds. Training code, model checkpoints and interactive examples are provided as open source software.
\end{abstract}

\section{Introduction}


What happens when a musical instrument reflects oneself, diffracts cultural forms, or reveals alien aesthetics of computation \cite{fazi_contingent_2018} in an intimate dance with the user? How do we work with creative AI when its behavior is unique to the situation at hand, and can only be drawn out via interaction with the body? In the Intelligent Instruments Lab, we consider musical performance and improvisation an excellent domain to study such questions. Our wider research program involves designing a system of diverse technical elements, from software to hardware, sensors to effectors to processors, which can be readily assembled into `intelligent instruments' for probing the embodied experience of AI \cite{langspil}. As part of this program, we designed a low-latency probabilistic sequence model for MIDI streams while entertaining future transfer to non-MIDI sequences such as sensor data. Because it is a protean and flexible backbone\footnote{https://en.wikipedia.org/wiki/Notochord} for embodied musical tasks, we have named this model \emph{Notochord}.

Notochord is an open-ended tool for MIDI processing, designed to maximise the space of possible interactions while making few assumptions about input device or the user’s creative interests. An instrument designer using Notochord can program fine-grained interventions into the generative process which a performer can interact with in real-time. For example, one can require that the next MIDI event in a performance have a pitch-class of C, or that it will occur no sooner than 100 milliseconds, or that it will be played on the snare drum with a velocity of 99, while Notochord selects its other attributes in context. 

Notochord is intended for exploration of creative AI in a real-time performance setting. At a low latency where the delay between action and response is imperceptible, an instrument may begin to feel more like an extension of the body than an external content production device. Yet many creative AI applications involve delays on the order of seconds or longer between action and result. With such a coarse rate of feedback, it is difficult for them to enter the body schema \cite{Klinke_2014} in the manner of a guitar string or paintbrush. It is well known how musical applications can have particularly demanding latency requirements \cite{mcpherson_action_sound_2016}, and to ensure that Notochord does not disrupt embodied interaction with instruments built on it, our design emphasizes low-latency processing of each input MIDI event.

In Sections \ref{background} and \ref{model}, we describe the theory and implementation of the Notochord probabilistic model. In Sections \ref{training} and \ref{results}, we report on an instance of Notochord trained on the Lakh MIDI dataset to sponge up 100,000 songs worth of ambient musicality. Examples of specific applications built on Notochord are given in Section \ref{applications}.

\section{Background} \label{background}

The present work falls within what Ji et al. \cite{ji_comprehensive_2020} call ``composing expressive performance'' and Oore et al. \cite{oore_this_2020} call ``direct performance generation'': modeling music at a symbolic level, but with the inclusion of performed timing and gesture. Specifically, we aim to model real-time musical performance or improvisation captured via MIDI while also incorporating information from preprogrammed MIDI files.



The typical approach to representing MIDI performance is to use a `text-like' representation,\footnote{Note that we do not discuss \emph{literal} text representations, like the ABC notation used in the folkrnn line of work \cite{hallstrom_jigs_2019}, as they represent scores but not performances.} which flattens all musical structure into a linear sequence of categorical variables, exploding the sub-parts of notes to discrete events from a unified vocabulary. This renders musical data similar to text data, allowing methods from language processing to be transferred to music. PerformanceRNN \citep{oore_this_2020} for example uses separate `velocity change', `time shift', and `pitch' (on or off) tokens to model single MIDI events and the time between them. The MMM model \cite{ens_mmm_2020} and the model of Simon et al \citep{simon_learning_2018} both add program change events to handle multiple instruments, concatenating MIDI tracks sequentially. The REMI representation used for Pop Music Transformer \cite{huang_pop_2020} adds tempo and chord events, and uses time signature-aware `position' and `bar' events instead of time shift. 

Text-like representations are extremely flexible, but since they take multiple sub-events to represent each MIDI note, efforts have been made to improve computational complexity by grouping them back together. The MuMIDI representation used for PopMAG \cite{ren_popmag_2020} has a similar vocabulary to REMI, but introduces a method of summing embeddings to reduce total sequence length.
The NoteTuples representation used for Transformer-NADE \cite{hawthorne_transformer} also groups note features into single timesteps, each note being a tuple of (coarse time, fine time, pitch, velocity, coarse duration, fine duration). In contrast to PopMAG, they fully model the internal structure of each note using NADE \cite{uria_neural_2016}.


The probability model implemented by Transformer-NADE \cite{hawthorne_transformer} (itself inspired by \citep{boulanger-lewandowski_modeling_2012}) is similar to ours in that it is an autoregressive model for composite events. Differences include that we treat note-offs separately; we use continuous time and velocity instead of coarse/fine; and we use an RNN-based architecture for low latency prediction. The authors allude to, but do not elaborate on, any-order note factorization and discretized mixture logistic distributions, ideas which we developed independently for Notochord in Section \ref{model}. Notochord also bears similarities to infilling methods like MMM \cite{ens_mmm_2020} and MusIAC \cite{guo_musiac_2022}, though the focus there is on assisted composition rather than low-latency performance.

Our applications are inspired by work including the DeepBach chorale generation system \cite{hadjeres_deepbach_2017}, which explores fine-grained interventions into a probabilistic model; Piano Genie \cite{donahue_piano_2019}, which constructs an ``intelligent interface'' for piano performance using an autoencoder with a bottleneck reducing the 88 piano keys to eight controller keys; and Mann's \cite{noauthor_ai_nodate} and Castro's \cite{castro_performing_2019} efforts to wrangle pre-trained Magenta models into an interactive environment.

\section{Notochord} \label{model}

Notochord is a deep autoregressive model for sequences of events. Its main distinguishing features are chosen to support low-latency interaction, including musical performance. First, it uses a causal and order-agnostic event representation. Causal, in the sense that no future information is included in an event: voices are interleaved rather than concatenated serially, and note-off events are used rather than note durations. Order-agnostic, meaning that velocity and time-skip need not be predicted before pitch: within an event, the sub-events can be predicted in any order, which supports a range of applications as described in Section \ref{applications}. Second, the architecture is designed with low, fixed latency in mind: we use a recurrent backbone rather than long convolutions or self-attention.

In section \ref{representation}, we describe the data representation used for Notochord. In \ref{autoregressive} we introduce Notochord's probability model at the coarsest level, in \ref{order_agnostic} and \ref{sub-events} with increasing granularity. Finally in section \ref{nn} we describe the underlying neural network function approximators.

\subsection{MIDI representation} \label{representation}

Notochord uses a shallowly hierarchical event-based representation similar to Transformer-NADE and PopMAG. In contrast to those works, we separate notes into on and off events rather than using duration. This allows for low-latency applications where note duration may not be known at the time of onset. Events have internal structure (pitch, time, velocity, instrument), which we model explicitly to support a wide range of interpretable interventions.

In this work, we limit the task to modeling streams of MIDI Note On and Note Off\footnote{For clarity, we capitalize events from the MIDI spec (``Note On'') but hyphenate when referring to events in the Notochord representation (``note-off'')} events (plus implicitly program change events). Other MIDI messages (Pitch Bend, Control Change, Aftertouch) are ignored for simplicity, though we hope to include them in future work. We model the set $ X = \{ x^1 \dots x^M \} $ of MIDI sequences $ x^i = \{ x^i_1 \dots x^i_{N_i} \} $, where each event $ x^i_j = (\Delta t^i_j, v^i_j, p^i_j, \alpha^i_j) $ is composed of several sub-events: a continuous inter-event time $\Delta t$, continuous velocity $v$, categorical pitch $p$, and categorical instrument identity $\alpha$. Our note-off events are always encoded as events with velocity zero, with any release velocities being ignored. (Release velocities are inconsistently present in MIDI data since many controllers and synthesizers do not support them, and instead use Note On with velocity 0 to represent Note Off, as we do).

In contrast to much previous work which represents time using quantized, tempo-relative units, we represent time in seconds, as a continuous quantity. This lets our model handle tempo changes and performances in free time or without requiring any predefined metric structure. As in live music performed without a metronome,  all metric structure is implicit in the sequence of events which make up a performance.

In our system, each event includes an instrument ID, in contrast to MIDI where each event has a channel from 1-16, and a separate Program Change event sets the instrument of a channel. This extends MIDI in the sense that we are not limited to 16 instruments at once; it has the limitation that we cannot represent multiple instances of the same instrument. For example, in our representation all 128 General MIDI instruments can sound simultaneously. However, it is not possible to have two ``tenor saxophone'' instruments playing the same pitch at once (though see Appendix \ref{data} for more on our instrument representation).

\subsection{Autoregressive factorization} \label{autoregressive}

Like many other deep generative models, Notochord is probabilistic. Given a stream of events, it assigns a numerical probability to that stream, and this is how it is trained: to maximize the probability assigned to actual streams in a dataset. In application, it can stochastically \emph{sample} new streams according to the probabilities it has learned. Furthermore, Notochord is designed to be used interactively: when we sample each event, we need to do it quickly, and we can assume that all past events are known but no future events are known. 

An autoregressive model uses exactly this strategy to model complicated objects (like long streams of events) in terms of simpler objects (single numbers). More formally, we factor a joint distribution of high-dimensional data points $P(x)$ into a product of simpler conditional distributions $\prod_i P(x_j | x_{< j})$. Such a model can be fit by maximizing the conditional probability of data with respect to model parameters $\theta$, resulting in the objective: $\max_\theta \sum_{i,j} \log P(x^i_j | x^i_{< j})$.

\subsection{Sub-event order} \label{order_agnostic}

Recall from Section \ref{representation} that each of our events $x^i_j$ is a tuple of multiple sub-events (instrument, pitch, time and velocity). These musical quantities are not statistically independent, even given all previous notes; the next pitch will depend on which instrument plays it, velocity will depend on whether the timing indicates a strong beat, and so on. In fact, we envision those dependencies as affordances for making meaningful interventions in the generation process. For example, suppose a user wants to constrain the next event to have a specific pitch of G\#3.
In this case, they would query the model as follows: ``given that the next event has a pitch of G\#3, which instrument will play it and when?''

Sub-event structure can also be modeled autoregressively: we might first sample the instrument, which would then condition the sampled pitch, then time, then velocity. That is, we would factor $ P(x_j | x_{<j}) = P(\alpha_j | x_{<j}) P(p_j | \alpha_j, x_{<j}) P(\Delta t_j | p_j, \alpha_j, x_{<j}) P(v_j | \Delta t_j, p_j, \alpha_j, x_{<j}) $

What if we want to intervene within an event? Suppose we want to insist that the next event be from the grand piano (i.e. General MIDI instrument 1). In that case, we simply fix $\alpha_j=1$ instead of sampling from $P(\alpha_j | x_{<j})$. But consider a different application: we want to enforce that the next event has velocity zero (i.e. is a note-off) but we want the model to decide which note to end and when. If we simply fix $v_j = 0$ instead of sampling $P(v_j | \Delta t_j, p_j, \alpha_j, x_{<j})$, our intervention will have no causal effect on the rest of the event, since the other parts were sampled first. The model might choose a note which is not even currently playing. The same logic applies to further applications: if we want the next note to be in a high register, we should fix it before sampling the instrument -- what if the model, naïve to our requirement, samples the bass?

To enable a user to query in \emph{any} desired order at inference time, our solution is to optimize over \emph{all} permutations of sub-event parts. During training, each sub-event prediction is conditioned upon a random subset of the other sub-events.

\subsection{Sub-event distributions} \label{sub-events}

With our events now broken into scalar sub-events (instrument, pitch, time, velocity), we can model each with a parametric probability distribution. Instrument and pitch are categorical variables, so we represent their conditional distributions in the typical way, with a vector of probabilities produced from the $\operatorname{softmax}$ function. 

Time and velocity, however, are continuous in our model (Section \ref{representation}). Since MIDI files can contain a variety of tick durations and ticks per beat, we dequantize time and convert to absolute times in seconds. While MIDI velocities take only 128 values, we choose to also dequantize velocity and treat it similarly to time with an eye toward future transfer learning to non-MIDI domains with finer dynamics (raw data from a piezoelectric sensor for example). A discretized mixture of logistics \cite{salimans_pixelcnnxx_2017} is used to model the values of velocity and time. We elaborate on this choice in Appendix \ref{dmol}.

\subsection{Neural network architecture} \label{nn}

\begin{figure}
  \centering
  \includegraphics[width=0.7\linewidth]{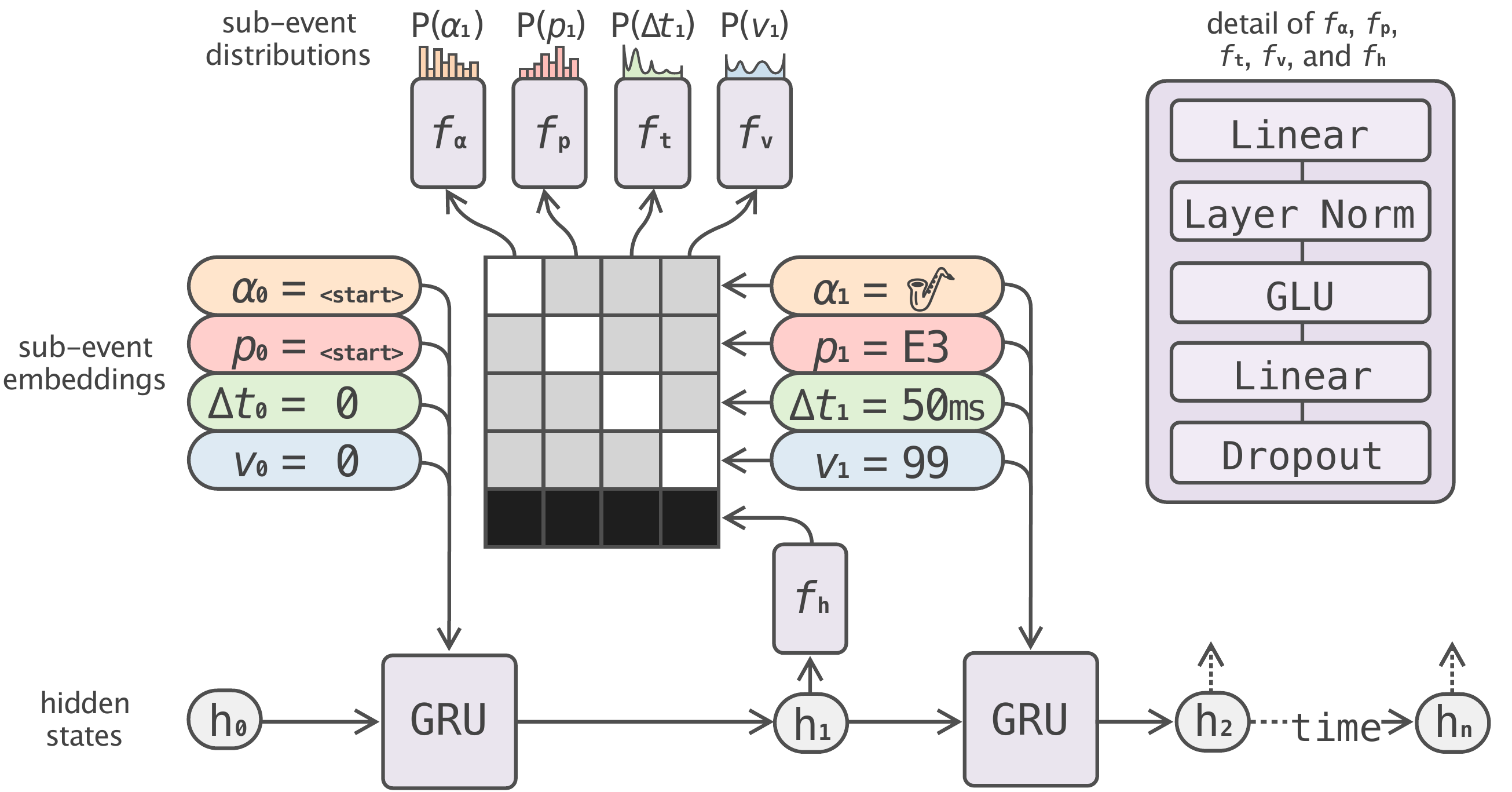}
  \caption{Architecture of the Notochord model at training time. Rectangular blocks are functions, long capsules are embedding vectors, and short capsules are hidden states. Each sub-event depends on previous events via a GRU, and also on a random subset of the other sub-events. Conditioning of each sub-event on other sub-events is achieved by simply adding their embeddings to the hidden state after passing it through an MLP $f_h$. The addition can be implemented in parallel as a batched matrix multiplication at training time. This is depicted with black cells indicating a one, gray cells a random binary value as proposed in Section \ref{order_agnostic}, and white cells a zero. A final MLP per sub-event maps the summed embeddings and hidden states to distribution parameters. MLP architecture is shown as an inset, top right.}
\label{fig:arch}
\end{figure} 

Thus far, we have described Notochord in the abstract as a probability model factored into conditional distributions. In this section, we describe the function approximators used to learn those conditional distributions from data.

To embed sub-events into vector space at the inputs to networks, we use a standard look-up table embedding for categorical variables. For continuous variables, we use a sinusoidal embedding, elaborated on in Appendix \ref{sines}

We implement the causal dependency between events using a gated recurrent unit (GRU) network \cite{cho_learning_2014}. This allows for low latency processing of single events at inference time compared to memory-less architectures for which computational cost scales with receptive field. Embeddings for sub-events are summed to produce a single input embedding per event, which is passed into the GRU to produce a hidden state $h_i$ which depends on all previous events $x_{<i}$.

Dependency between concurrent sub-events is achieved by summing the embeddings for conditioning sub-events into the hidden state. Recall that we optimize over all permutations of sub-event orders (Section \ref{order_agnostic}); to do so, we randomly select a subset of other sub-events to condition each target sub-event. The conditioned hidden state is passed through a multilayer perceptron (MLP) for each sub-event modality to produce distribution parameters for the sub-event, i.e. logits in the case of categorical modalities (instrument, pitch), and mixture weights, locations and scales for the continuous modalities (time, velocity). For example, to compute probability of the pitch of the $i$th event given the $i$th velocity and time-difference, as well as all previous events, we have: 
$$P( p_i | x_{<i}, v_i, \Delta t_i ) = \operatorname{softmax}( f_p(f_h(h_i) + v_i + \Delta t_i))$$
Where $h_i$ is the $i$th GRU hidden state, $f_p$ and $f_h$ are the MLPs for pitch and hidden state, and $v_i$ and $\Delta t_i$ are here the embeddings for velocity and time-difference.

All MLPs $f_\alpha, f_p, f_{\Delta t}, f_v$, and $f_h$ have the same architecture using dropout \cite{hinton_improving_2012}, layer normalization \cite{ba_layer_2016}, and gated linear unit (GLU) activations \cite{dauphin_language_2017}. We also fit a linear end-of-sequence predictor conditioned on the hidden state. The Notochord architecture is depicted in Figure \ref{fig:arch}.

\section{Training} \label{training}

Notochord is implemented in PyTorch \cite{pytorch} using standard layers, plus our own implementation of the discretized mixture of logistics (Appendix \ref{dmol}), with reference to that of \cite{salimans_pixelcnnxx_2017}.

Notochord is fit to the Lakh MIDI dataset (LMD) \cite{lakh}\cite{raffel_learning_based_2016}. LMD is not specifically a performance dataset, containing mostly programmed songs. We nevertheless chose to develop Notochord using the LMD as it is large, noisy, and diverse, to emphasize robustness and flexibility in the design. We leave integration of more narrow but performance-oriented datasets like Groove MIDI \cite{groove2019} and GiantMIDI-Piano \cite{kong_giantmidi_piano_2022} to future work. Details of data processing and augmentation are given in Appendix \ref{data}, and optimization in Appendix \ref{optimization}.

\section{Results} \label{results}

\begin{figure}
  \centering
  \includegraphics[width=0.5\linewidth]{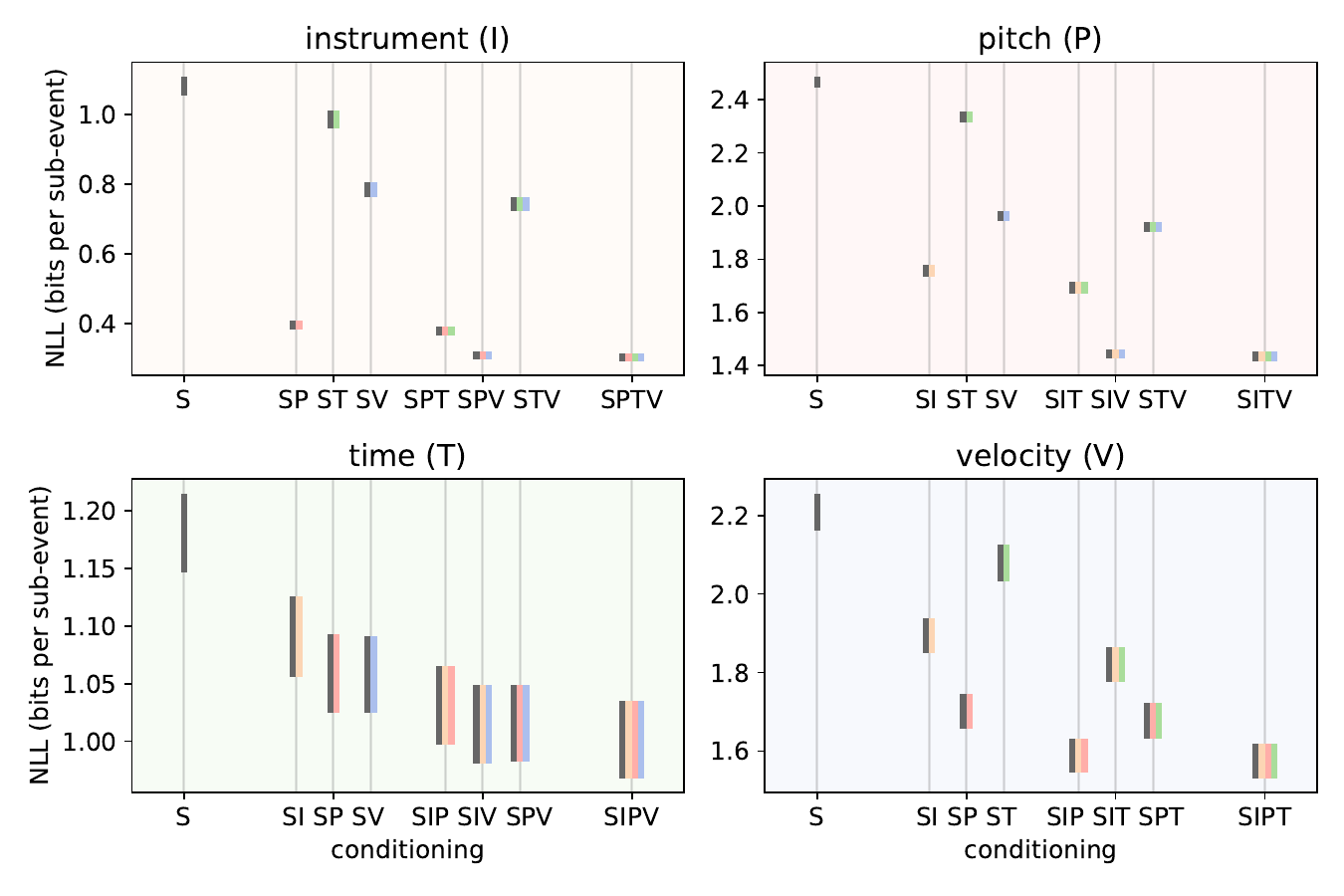}
  \raisebox{0.07\height}{
    \includegraphics[width=0.38\linewidth]{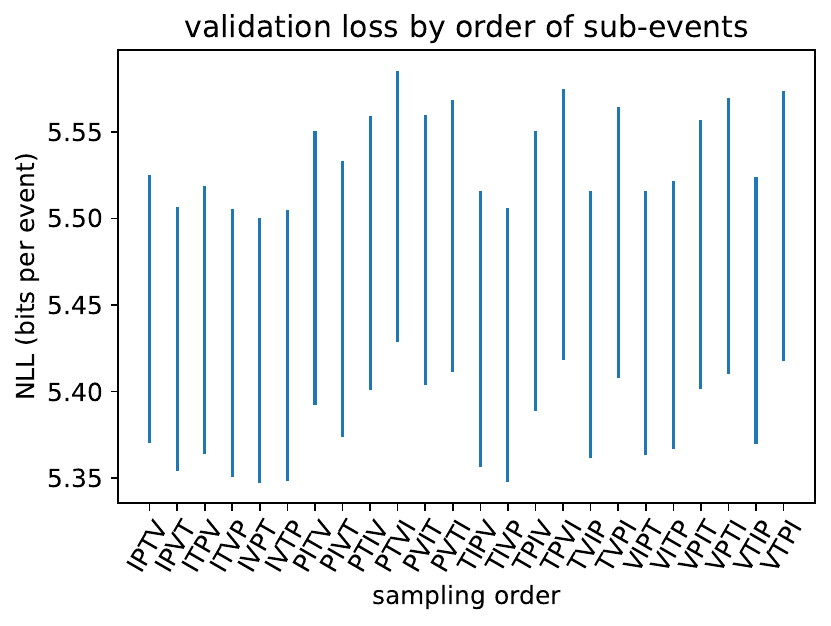}}
  \caption{Bootstrap 99\% confidence intervals for negative log likelihoods (NLL) computed over the validation set (lower is better). On the left, NLL is broken out by sub-event modality (instrument, pitch, time, velocity) and by which other sub-events each is conditioned on. In the leftmost position of each subplot, the sub-event is conditioned only on previous events via hidden state (S) and then from left to right on larger combinations of other sub-events. On the right, total NLL per event is reported for every permutation of sub-event order.}
\label{fig:nll}
\end{figure}

In Figure \ref{fig:nll}, we investigate the efficacy of any-order event factorization (Section \ref{order_agnostic}). We can see that the negative log-likelihood consistently decreases as more information is available. This indicates that the model is successfully conditioning on all available information, and we can expect sub-event interventions to be meaningful. The effect of sampling order on total likelihood is small, but sampling instrument earlier seems to be advantageous.

Cursory timing on a MacBook Pro with Intel Core i7 7700HQ processor gives about 6ms to feed an event to the model and about 3ms to sample the next full event. In future work, we hope to improve on this using pruning and quantization.

\begin{figure}
  \includegraphics[width=0.5\linewidth, trim=0 2cm 0 3cm]{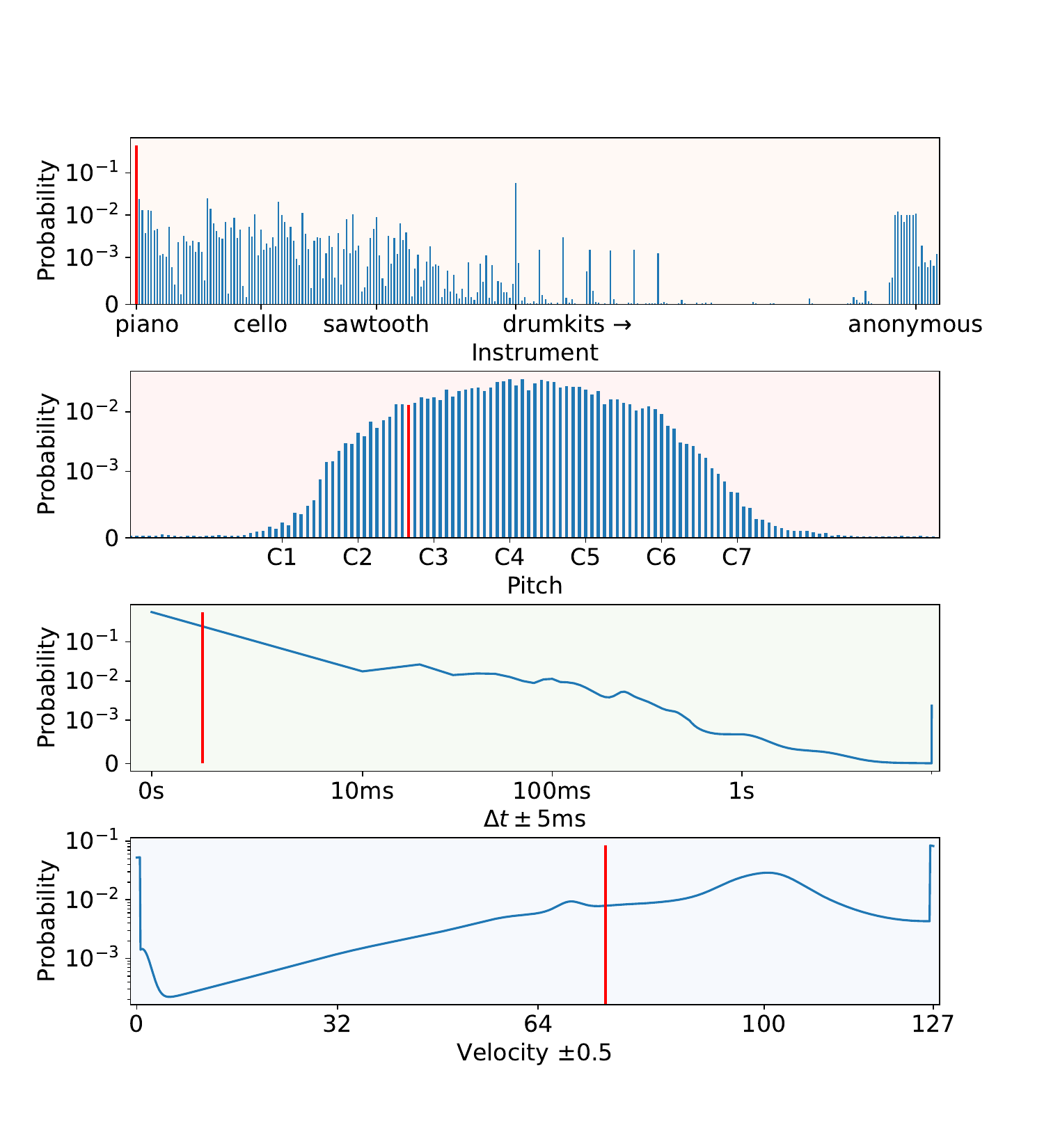}
  \includegraphics[width=0.5\linewidth, trim=0 2cm 0 3cm]{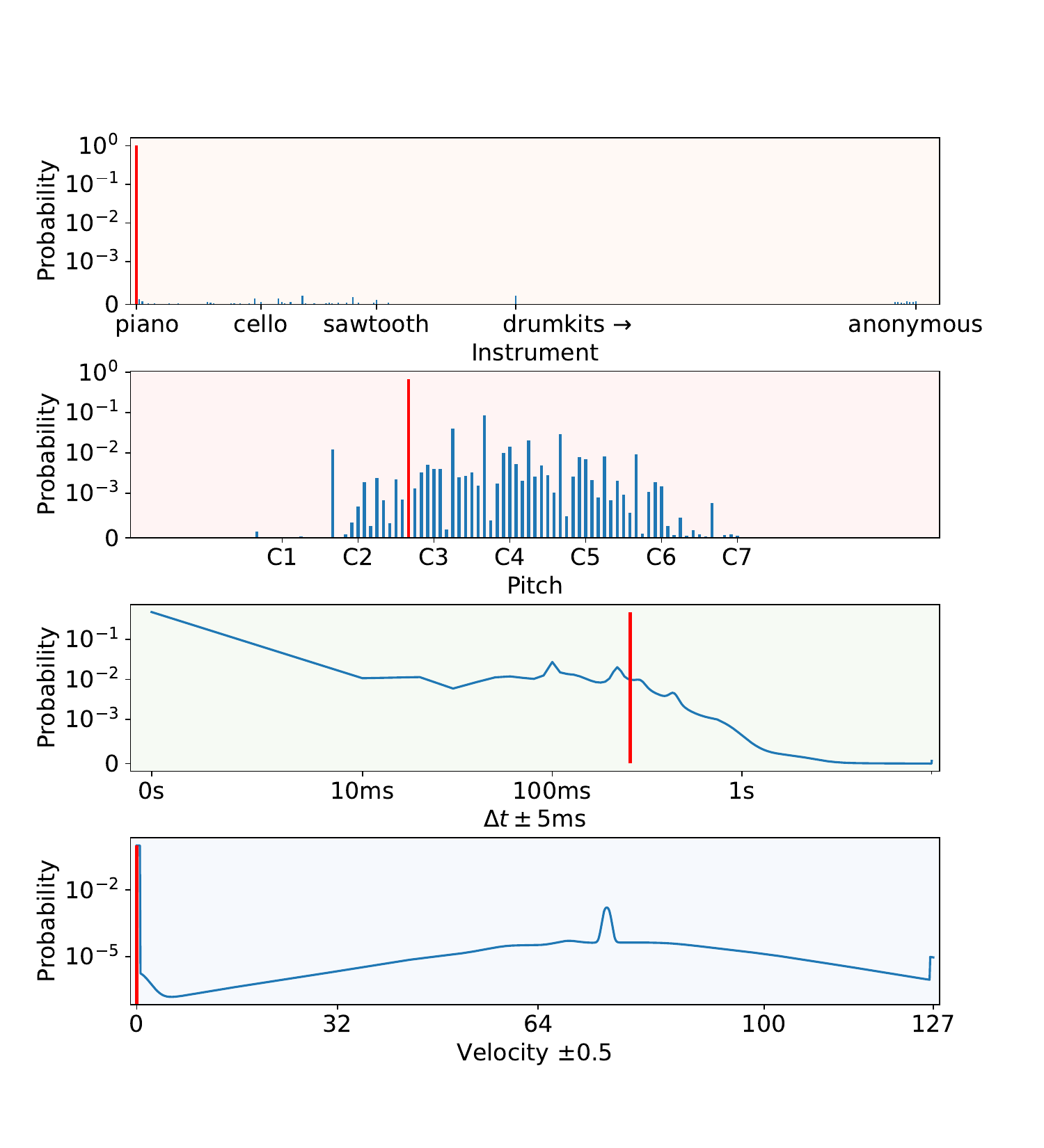}
  \caption{A sequence of conditional distributions (Section \ref{sub-events}) from sampling the model. Sub-events are ordered from top to bottom, then events left to right; red lines indicate sampled values. In this example, the discrete distribution over instrument (orange, top left) is sampled first, then pitch (pink), then the mixture density over time (green), and velocity (blue). Sampling continues in the right column, beginning again with instrument for the second event. Note how the initially higher entropy of the instrument distribution (top left) collapses to a very high probability of sampling the same instrument again (top right); and how the velocity value sampled first (bottom left) becomes a more likely value for the second sample (bottom right)}
\label{fig:distributions}
\end{figure}

Notochord does not compete with the state of the art when used for coherent music generation, because it aims for low-latency interactivity and enforces few assumptions about musical structure. For this work, we focused on bridging the gap to very low latency and the diversity of applications it can enable, leaving a thorough comparison of neural network architectures to future work. Figure \ref{fig:distributions} illustrates a series of sub-event distributions as the model is sampled.

\section{Applications} \label{applications}

In this section, we illustrate the potential of Notochord with several preliminary applications. These are implemented to varying degrees of completion in our open-source repository.\footnote{https://github.com/Intelligent-Instruments-Lab/iil-python-tools/tree/master/examples/notochord} 
A Notochord-based application typically consists of three parts:
\begin{itemize}
    \item The Notochord server running in Python, with methods to\verb_ feed_ each MIDI event to Notochord as it happens and to \verb_query_ for predicted future events over Open Sound Control (OSC).
    \item A front-end scheduler defining the application, which handles MIDI inputs and communicates with Notochord. It determines what to feed, how to query and what to do with the responses. Our example front-ends are built with SuperCollider \cite{noauthor_supercollider_nodate}.
    \item A synthesizer which converts the MIDI streams to sound. This might be hardware, a DAW, or a General MIDI implementation such as fluidsynth \cite{fluidsynth}.
\end{itemize}




\subsection{Steerable generation}
We can get a first listen to how Notochord behaves by sampling streams from events one at a time and sending them to a MIDI synthesizer. Streams sampled from Notochord are rarely convincing imitations of the data, but they have a certain ``vaporwave fantasia'' charm, as one observer put it; Notochord can be a prism for diffracting General-MIDI culture into weird retrofuturistic skeins.

\begin{figure} \label{fig:pianoroll}
  \centering
  \includegraphics[width=0.7\linewidth]{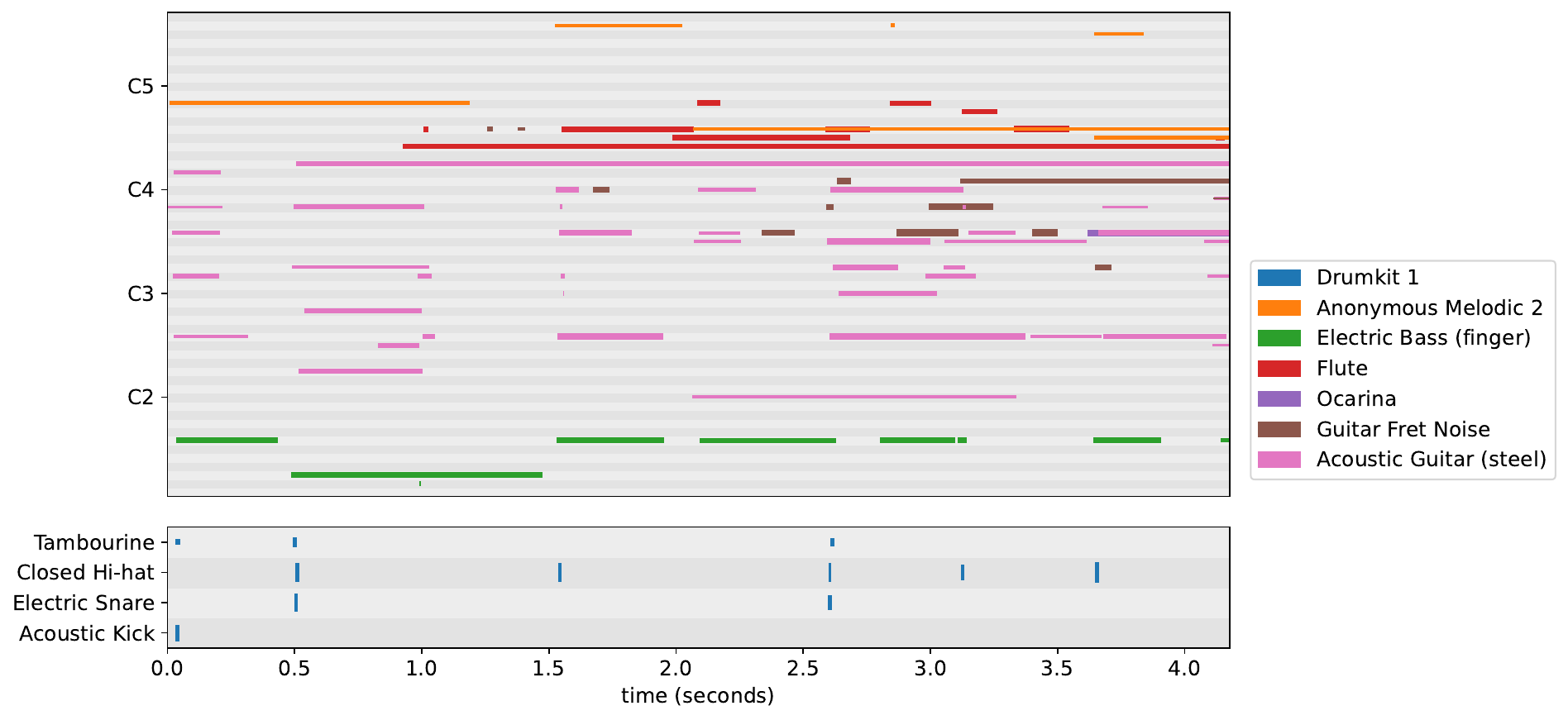}
  \caption{Piano-roll visualization of event streams generated by sampling Notochord. We encourage diversity by sampling the instrument of the first event uniformly from the General MIDI instruments instead of using the model prior, which like the LMD is heavily biased toward instrument 1 (see Figure \ref{fig:distributions}).}
\end{figure}

Generation can be `steered' by manipulating the predictive distribution before sampling each sub-event. For example, timing can be truncated to control event density; or pitch can be limited to certain register; or a specific set of instruments can be selected. We can stop when the model predicts a sequence end, or keep sampling and see where else it goes. Figure \ref{fig:pianoroll} illustrates an event stream sampled from Notochord.


\subsection{Auto-pitch and neural harmonizer} \label{autopitch}
Because Notochord can handle the sub-parts of MIDI events in any order (Section \ref{order_agnostic}), it can be used to `fill in the blanks'.
To build an `auto-pitch' instrument, we take timing, velocity and instrument identity from a MIDI controller or other source and query only pitches from Notochord. A player can drum on a single pad controller, for example, and Notochord will generate a melody fit to the rhythm and intensity of the performance in real-time.



\begin{figure} \label{fig:harmonizer}
  \centering
  \includegraphics[width=0.7\linewidth]{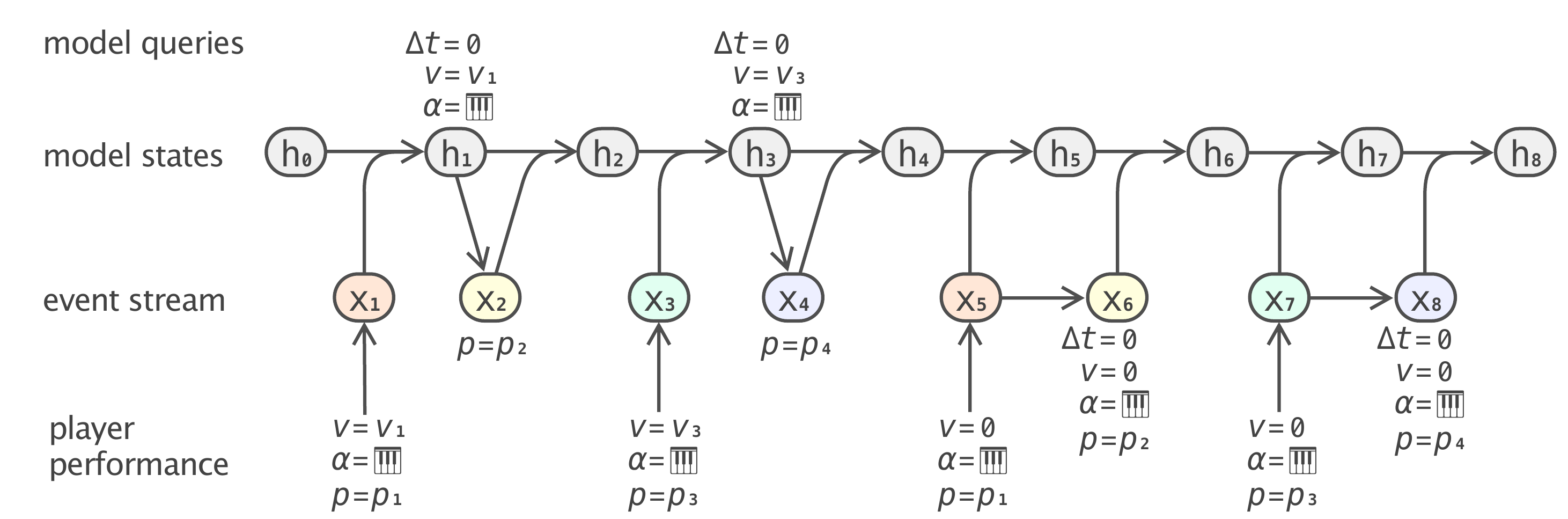}
  \caption{Implementation of the neural harmonizer. Input events from a MIDI controller are in the bottom row. At the top is a sequence of model states annotated with queries for each harmonizing pitch. The combined stream of events from the player, Notochord, and the scheduler appear in the middle. In this example, the player strikes two notes before releasing each of them. The scheduler tracks which harmonizing pitches are associated with which performed pitches in order to generate matching note-offs.}
\end{figure}

In a slightly different scenario, we can take complete incoming MIDI events and use them to query additional events from Notochord. If we answer every note-on event from the player with a sample constrained to have $\Delta t = 0, v > 0$, we have an `intelligent' harmonizer which is sensitive to the entire performance so far. Figure \ref{fig:harmonizer} illustrates the interaction between performer, scheduler and Notochord to achieve this.

\subsection{Live coding with TidalCycles}

TidalCycles (Tidal for short) is a popular language for live coding of pattern created by Alex McLean \cite{mclean2010tidal}.
We created an OSC target allowing Tidal to communicate with Notochord via SuperCollider.
In this case, the user specifies all timing and note-offs implicitly via pattern structure in Tidal. Instrument, pitch and velocity for note-ons can be completed by Notochord in a fine-grained manner (Listing \ref{lst:tidal_choosepitch}).

\begin{lstlisting}[
  xleftmargin=.2\textwidth,
  basicstyle=\ttfamily,
  columns=fixed,
  language=haskell,
  caption={
    An example Tidal pattern where Notochord `chooses' the pitch when it receives a -1.
    The instrument parameter is patterned as clavinet, marimba, organ, jazz guitar (according to General MIDI).
    The velocity pattern follows a sinusoid between values of 70-120.},
  label={lst:tidal_choosepitch}
]
p "choosepitch"
  $ ncinst  " 8 [ 13*3     [17 27]]"
  # ncpitch "-1 [[60 -1 60] -1    ]"
  # ncvel (range 120 70 $ sine)      
\end{lstlisting}

\subsection{Machine improvisation}
To improvise with Notochord as a partner we reserve certain instruments for any non-Notochord players and zero the probability of choosing those instruments when querying Notochord. Every event, player- or model- generated, causes a query for a new Notochord-generated event, which gets scheduled to occur after its $\Delta t$. If another player-generated event occurs first, the scheduled event is canceled and a new prediction is queried for. In other words, if Notochord `plans` to play something but a player goes first, it will `listen` and reconsider.

Being fit largely to MIDI arrangements of songs, the pre-trained Notochord is not the most considerate improvising partner! However, we can imagine building more interesting bespoke improvisers on top of Notochord's implicit musical `knowledge' and notion of `surprise', or fine-tuning on bespoke MIDI datasets.

\subsection{Likelihood-based interfaces}
Each of the previous examples involves random sampling from the Notochord model. Instead, we can use the probability scores it returns to design new musical interfaces. For example, a player might choose pitches ordered by their likelihood according to the model rather than by fundamental frequency as on a traditional keyboard. Pictured in Figure \ref{fig:linn} is an interface built on the Linnstrument \cite{noauthor_linnstrument_nodate} to choose pitches by likelihood.

\begin{figure}
  \centering
  \includegraphics[width=0.33\linewidth]{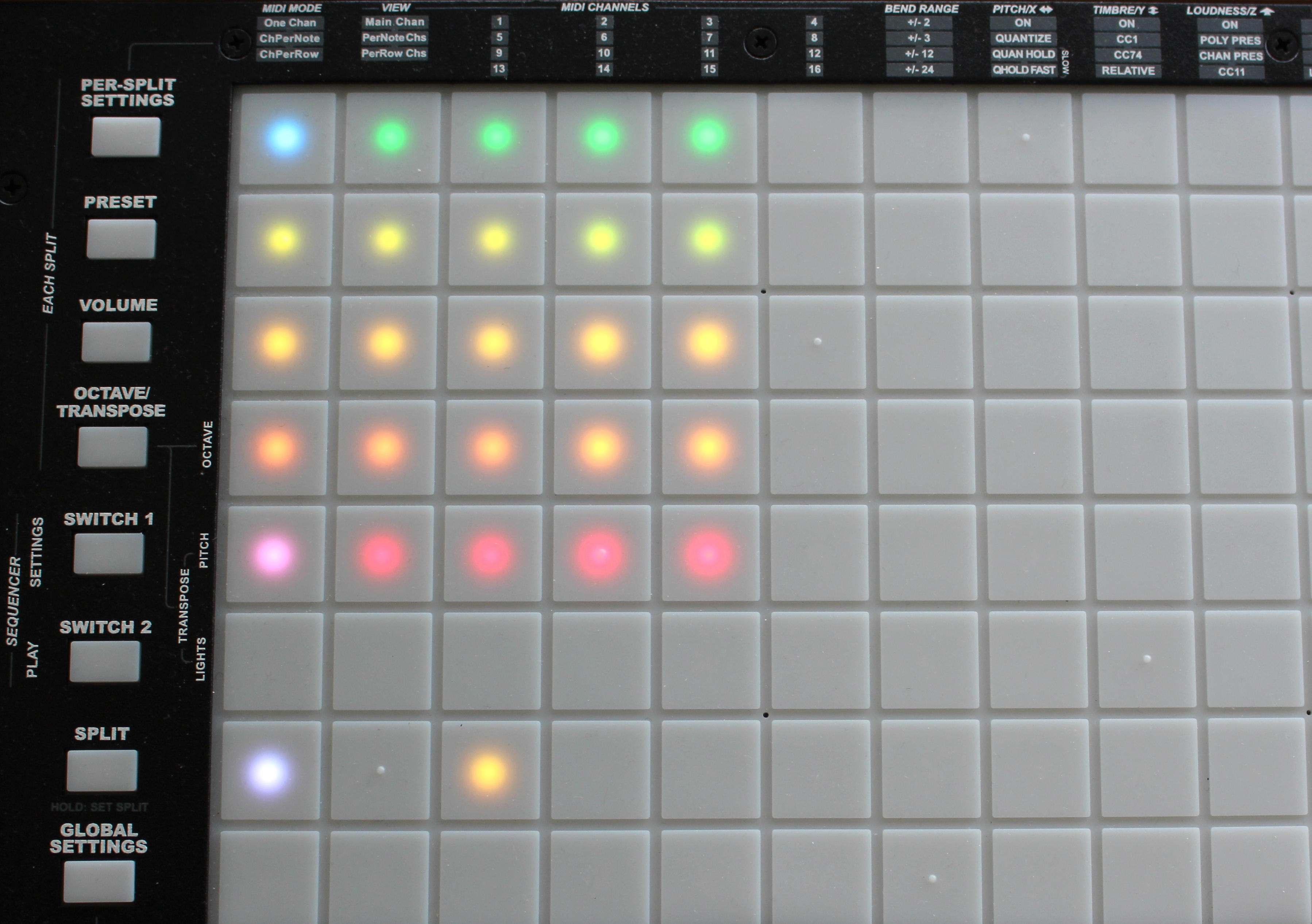}
  \caption{Linnstrument interface for likelihood-based auto-pitch. Timing and velocity come directly from the player via the Linnstrument pads; Notochord creates a dynamic mapping from the pad coordinate to pitch. Here the main grid allow selection of pitches by relative likelihood from the single most likely pitch (cyan) to the least (magenta). The single white pad samples pitch at random from the model distribution (Section \ref{autopitch}) and the yellow pad resets the model to its initial state.}
\end{figure} \label{fig:linn}

Rather than querying Notochord for predictions at all, we could measure the likelihood (degree of `surprise') that Notochord ascribes to events and use it for something else, like modulating a synth parameter.

\section{Conclusion}

This paper described Notochord, a new model for MIDI sequences which builds on previous deep learning-based methods, but with new affordances. Namely, it can respond (perceptibly) instantaneously in a real-time setting while also enabling fine-grained interpretable interventions, qualities which facilitate research into the embodied experience of machine intelligence in musical instruments. We concluded by sketching some early applications to highlight Notochord's flexibility. Code and model checkpoints for Notochord are provided as open-source software in the hope that others will experiment with it.

\begin{ack}
Thanks to Rui Guo and Davíð Brynjar Franzson for their valuable comments, and to Krish Ravindranath for discussions and code contributions.

The Intelligent Instruments project (INTENT) is funded by the European Research Council (ERC) under the European Union’s Horizon 2020 research and innovation programme (Grant agreement No. 101001848).

INTENT is also supported by an NVIDIA hardware grant of two A5000 GPUs. 

\end{ack}

\bibliographystyle{plainnat}  
\bibliography{aimc2022}

\appendix
\section{Time and velocity distributions} \label{dmol}
We use a discretized mixture of logistics \cite{salimans_pixelcnnxx_2017} to model velocity and time sub-events, which allows us to compare the probabilities of continuous quantities (being within some interval) to those of discrete quantities. Thus our generative model treats velocity and time as continuous quantities which have been quantized, which is the case when dealing with MIDI files which are captured performances. But it can also handle intrinsically discrete data, like MIDI sequenced on a piano-roll. We use a resolution of $r=1$ for velocity (which remains valued from 0-127), and of $r=10$ms for time, to limit the sensitivity of the model to tiny differences which are an artifact of quantization in MIDI data. At training time, we model the probability that data is within $\pm \frac{r}{2}$ as a difference of values on the cumulative distribution function (CDF). At inference time, the learned CDF defines a probability density from which we sample continuous values. For further details, consult Salimans et al. \cite{salimans_pixelcnnxx_2017}.

The use of a mixture distribution for time is inspired by the discrete character of rhythmic intervals -- there is typically a finite set of rhythmic intervals which make sense musically (quarter note, triplet), but within each there is room for variation in the fine timing (groove). Selecting a mixture component can be viewed as `sampling the rhythm`, and then sampling from it `samples the timing'. It is common to modify parametric distributions before sampling as a way of tweaking the results; `temperature' sampling adjusts the balance of high- and low-probability outcomes. We can separately modulate `rhythm temperature' and `timing temperature' by altering the mixture weights and component scales, respectively.

\section{Sinusoidal embedding} \label{sines}

To embed continuous scalar inputs to our model, we use a vector of sinusoids followed by a linear projection, inspired by the Fourier features of Tancik et al. \cite{tancik_fourier_2020}. Sinusoids are logarithmically spaced by wavelength for time, and linearly for velocity. For example, a scalar velocity $v$ is first mapped to a vector $v_s = \sin(2 \pi f_1 v), \sin(2 \pi f_2 v), \ldots \sin(2 \pi f_N v) $ and then to the shared embedding space with $v_e = W_v v_s + b_v$, where the weight matrix $W_v$ and bias vector $b_v$ are learnable parameters but the frequencies $f$ are fixed. For further details, see our open source implementation.
\footnote{https://github.com/Intelligent-Instruments-Lab/iil-python-tools/tree/master/notochord}

\section{Data processing and augmentation} \label{data}

Instruments are extracted from each MIDI file using \texttt{pretty\_midi} \cite{raffel_intuitive_2014}, which we rely on to convert tempo-relative timing to seconds and interpret MIDI Note Off events. We further trim any notes sharing the same instrument and pitch (`the same key of the piano') to not overlap and leave at least 1ms between note-off and note-on, so they won't be transposed when we add a small temporal jitter later.

At training time, we use data augmentation to forestall overfitting and mitigate biases in the data toward default keys, velocities and tempi. We apply random global tempo change of $\pm 10\%$, a random transposition of $\pm 5$ semitones, and a random velocity curve with an exponent log-normally distributed with $\mu = 1, \sigma = \frac{1}{3}$. We also apply a temporal jitter of $\pm 1$ millisecond independently to each event, which has the effect of randomizing the order in which concurrent events appear to the model, while remaining imperceptible. We dequantize velocity as discussed in Section \ref{sub-events}, but without disturbing the extreme values of 0 and 127, since hard zeros have the special meaning of note-off. 

Finally, all instruments are recombined into one temporally-ordered stream with each event carrying the instrument number. Melodic instruments use the General MIDI standard 1-128, while drums are mapped to numbers 129-256. General MIDI uses a specific MIDI channel to identify drums, which share program and pitch numbers with the melodic instruments; since we do not use channels, drums are mapped to a distinct range of instrument IDs.

We also randomly map instruments to eight additional `anonymous' melodic and drum identities with a probability of $10 \%$ per instrument. This requires the model to infer instrument identities for making good predictions in these cases. At inference time, anonymous instruments can then be used in applications where bias toward a particular General MIDI instrument is undesirable.

\section{Optimization details} \label{optimization}
 
We train Notochord on LMD with minibatches of size 32, with a batch length starting at 32 events and increasing by 1 each time through the dataset. Batch size is increased to 64 at after ten billion events, then to 128 after another five billion. The AdamW optimizer \cite{loshchilov_decoupled_2019} is used with hyperparameters of $ ( \gamma = \num{1e-4}, \beta_1 = 0.9, \beta_2 = 0.999, \epsilon = \num{1e-8}, \lambda = 0.01 )$.

We use dropout with $p=0.1$ in our MLPs. Gradient L2 norm is clipped to a maximum of $1$. Masks for the sub-event dependencies are sampled independently for every batch item and time step. Lacking the resources for exhaustive hyperparameter sweeps, we relied on preliminary experiments, hunches and best practices.

We train for up to twenty billion total events, which takes about 80 hours on a single A4000 GPU, and reaches a batch length of about 500 events. A validation set of $5\%$ of the data was used to monitor for overfitting, but we observed none (validation loss maintained a downward trend).

\end{document}